\documentclass[11pt, reqno]{amsart}
\usepackage{geometry,amssymb,amsmath,amsthm,graphicx,hyperref}
\usepackage{xcolor,tikz}
\usepackage[comma, sort&compress]{natbib}

\newtheorem{theorem}{Theorem}[section]

\newtheorem{defin}[theorem]{Definition}
\newtheorem{remark}[theorem]{Remark}

\newcommand{\one}{\mathbf{1}}
\newcommand{\bd}{\mathbf{d}}
\newcommand{\br}{\mathbf{r}}
\newcommand{\bw}{\mathbf{w}}
\newcommand{\bu}{\mathbf{u}}
\newcommand{\bv}{\mathbf{v}}
\newcommand{\bx}{\mathbf{x}}
\newcommand{\by}{\mathbf{y}}
\newcommand{\R}{\mathbb{R}}
\newcommand{\scalar}[2]{\langle #1,#2\rangle}

\title[Centrality Paradox]{The Generalized Friendship Paradox for Spectral Centralities}

\author[R.~S.~Hazra]{Rajat Subhra Hazra}
 \address[RSH]{ Mathematical Institute, University of Leiden, Einsteinweg 55,
2333 CC Leiden, The Netherlands.}
\email{r.s.hazra@math.leidenuniv.nl}
\author[E. Verbitskiy]{Evgeny Verbitskiy}
\address[EV]{Mathematical Institute, University of Leiden,  Einsteinweg 55,
2333 CC Leiden, The Netherlands \textit{and} Korteweg–de Vries Institute for Mathematics, University of Amsterdam, Postbus 94248,
1090GE Amsterdam, The Netherlands}
\email{evgeny@math.leidenuniv.nl, e.a.verbitskiy2@uva.nl}
\date{\today}

\begin{document}
\maketitle

\begin{abstract}
We revisit the classical friendship paradox which states that on an average one’s friends have at least as many friends as oneself and generalize it to a variety of network centrality indices. For a broad class of spectral centralities on connected undirected graphs—degree, eigenvector centrality, walk counts, Katz centrality and PageRank, we show that the average centrality of a node’s neighbours always exceeds the global average centrality. We further prove an analogous result for PageRank on strongly connected directed graphs.  For
degree, this recovers the classical friendship paradox, while for the other centralities it yields new instances of what we call the centrality paradox. We also compare our neighbour-averaged formulation with edge-sampled versions studied previously in the literature.

\end{abstract}
\section{Introduction}\label{sec:intro}

The \emph{friendship paradox} was first noted by \citet{Feld1991} and it states that on average, your friends have more friends than you do. Although noted as a paradox, this can be easily established as a fact on any general graph with minimal assumptions. The phenomenon arises from a sampling bias, that is,  high-degree nodes appear more frequently in local neighborhoods, increasing the average degree seen by a typical individual. {The paradox is more than a curiosity: it shows a systematic sampling bias that affects how we perceive popularity, influence and norms in networks. A line of
work on the generalized friendship paradox studies attributes other than degree
that display the same effect. \citet{eom2014generalized}, for example, analyse co-authorship
networks from \emph{Physical Review} journals and Google Scholar profiles and
find that, on average, an author’s co-authors have more collaborations,
publications and citations than the author. On Twitter, most users follow
accounts that are more active and more popular than themselves: their friends
tend to share and retweet viral content more often, and the vast majority of
users have fewer followers than the people they follow~(\citet{hodas2013friendship}). From a modelling
perspective, Cantwell et al.~\cite{cantwell2021friendship} relate such generalized friendship paradoxes to positive correlation between the attribute of interest and a functional of degrees. There are other
studies link these sampling effects to perception biases (\citet{kumar2024friendship, wilson2010using}). In \citet{jackson2019friendship} the friendship paradox is linked to systematic biases in perception and to the spread of opinions. The key idea is that social norms are shaped by how individuals perceive the behaviours of those around them. For example, studies have found that people are more likely to smoke when they have acquaintances who smoke (\citet{christakis2008collective}).
}

The friendship paradox asserts that in any finite, undirected connected graph $G=(V,E)$,
\[
\frac{1}{|V|}\sum_{i\in V}\Bigl(\,\frac{1}{d_i}\sum_{j\sim i} d_j\Bigr)\ge \frac{1}{|V|}\sum_{i\in V} d_i 
\]
where $d_i$ denotes the degree of node $i$ and $j\sim i$ indicates adjacency.  
{Equivalently, if we first choose a node $i$ uniformly at random and then pick one of its neighbours uniformly at random, the neighbour’s degree dominates $d_i$. This result has been proved in a few different ways in the literature; see, for example, \citet{Feld1991, Piet:book, van2024random, cantwell2021friendship}. In Section~\ref{sec:examples} we include a short variational proof, phrased in terms of the random-walk matrix $C = D^{-1}A$ and a Perron--Frobenius inequality.
}

The degree of a node in a graph is itself a centrality index. Centrality indices assign a numerical score to each node in a network to capture its relative importance; common examples include degree centrality, PageRank, eigenvector centrality, Katz centrality, and betweenness centrality, each highlighting a different facet of  node’s influence. {It is therefore natural to ask for which centrality indices a friendship-paradox type inequality holds, that is, for which indices the average centrality of a node’s neighbours exceeds its own score.} In recent years, such generalizations have been suggested. For example, \citet{higham2019centrality} analyzes a version of the centrality paradox in which one chooses a random edge (friendship) and then a random endpoint of that edge, and compares the average centrality of those sampled friends to the global average.  \citet{cantwell2021friendship} showed that if you replace “degree” by any nonnegative attribute $x_i$ on each node $i$, then the difference between the average of $x$ over randomly sampled neighbors and the global average of $x$ is governed by the covariance between $x$ and a functional of the degree. In other words, whether your friends look “larger” on attribute $x$ than the average person depends precisely on how strongly $x$ is correlated with degree.

In this article we take a complementary perspective: rather than sampling an edge and one of its endpoints, we reformulate the paradox by averaging each node’s centrality over its neighbors. For $\br:V\to \R_{\ge0}$ a centrality index we compare the mean of $\{\br(j): j\sim i\}$ for each $i$ against the overall mean of $\br$. With this “neighbor‐averaging” definition in place, we then prove that the resulting centrality paradox holds not only for degree, but also for eigenvector centrality, walk counts (powers of adjacency matrix), Katz centrality, and even PageRank on strongly connected directed graphs.

\subsection*{\bf Main contributions and outline of the paper.}
To place our work within the existing literature, we summarize our main contributions and also provide an outline of the paper. After this introduction, in Section~\ref{sec:paradox-def} we set up notation for an undirected graph \(G=(V,E)\) and introduce the notion of centrality paradox and define five prototypical spectral centrality indices (degree, walk counts, eigenvector, Katz and PageRank) that will be studied.
\begin{itemize}
\item[(a)] We formalize a neighbour-averaged version of the generalized friendship paradox for a centrality vector \(\br\), expressed as the inequality
\(\langle \mathbf{1}, C \br\rangle \geq \langle \mathbf{1}, \br\rangle\), where \(C = D^{-1}A\). This formulation is distinct from the edge-sampled versions
studied in, for example, \citet{eom2014generalized} and \citet{higham2019centrality}.  In Section~\ref{sec:comparison} we prove the centrality paradox for each of the above examples, showing that on any connected graph the mean neighbor-centrality strictly exceeds the node’s own score.  

\item[(b)] We compare two
natural “friends’ averages” of centrality—node-wise neighbour averaging and
edge-based sampling—and show by examples that in general neither dominates the
other.  Section~\ref{sec:comparison} compares these two different friend-averaging statistics. 

\item[(c)] Finally, in Section~\ref{sec:discussion} we discuss limitations of our approach (in particular for distance-based measures), connections to sparse random-graph models via local weak limits, and promising directions for extending these results to more global or tree-like settings.

\end{itemize}

\section{Centrality Paradox}\label{sec:paradox-def}
Throughout Sections~2--4 we work with finite, simple, connected, undirected
graphs. Directed graphs enter only in Section~\ref{subsec:pagerank}, where we consider PageRank.
\subsection{Matrices and eigenvectors}
Let $G=(V,E)$ be a connected undirected graph (without self-loops) with $|V|=n$ nodes and $|E|=m$ edges. We use “node” and “vertex” interchangeably, and likewise “edge” and
“link,” and “graph” and “network". We label $V=\{1,\dots,n\}$.  Write \(A\) for its adjacency matrix, so \(A_{ij} \in \{0,1\}\) indicates whether
\(i\) and \(j\) are adjacent, and \(A\) is symmetric with zeros on the diagonal.
For simplicity we work with simple graphs, but all the arguments extend
verbatim to weighted undirected networks with a symmetric nonnegative
adjacency matrix. 
The degree of node $i$ is 
\[
d_i \;=\; \sum_{j=1}^n A_{ij}, 
\quad
\bd = (d_1,\dots,d_n)^\top,\quad D=\mathrm{diag}(d_1,\dots,d_n).
\]
Since $G$ is undirected and connected, $A$ is irreducible and nonnegative; its spectral radius $\rho(A)=\lambda_1>0$ is a simple eigenvalue with associated left/right Perron eigenvectors $\bv=\bu>0$ (we normalize $\|\bu\|_1=1$).  Define the transition matrix for the simple random walk on the graph by
\begin{equation}\label{eq:Cmatrix}
C\;=\;D^{-1}A\,,
\end{equation}
which is row‐stochastic ($C\one = \one$) and irreducible.  Thus $1$ is its Perron eigenvalue of $C$ with right eigenvector $\one>0$, and its unique positive left eigenvector $\bw^\top$ satisfies $\bw^\top C=\bw^\top$, $\bw^\top\one=1$.  In fact $\bw = d^\top / \|d\|_1$, since
\[
\bd^\top C \;=\;\bd^\top D^{-1}A 
\;=\;\one^\top A 
\;=\;\bd^\top.
\]

\subsection{Centrality indices}

A centrality index $\br: V\to [0,+\infty)$ is a non-negative function on vertices/nodes; we view $\br$ as a non-negative column  vector in $\R^n$. It assigns to each node a non‐negative score reflecting its “importance” within the network structure, by aggregating simple nodal statistics like counts of neighbours or walks, through a weighting function to capture how influence or connectivity decays with distance. We refer to the survey articles \citet{saxena2020centrality, shvydun2025zoo} and to the monograph of
\citet{avrachenkov2022statistical} for detailed classifications and
comparisons of centrality indices. Here we focus on the following important
examples.
\begin{enumerate}
\item {\bf Degree centrality:} Degree centrality is the most straightforward way to study a node’s importance: it simply counts how many connections the node has. In an undirected network, this equals the total number of edges incident on the node. In a directed network, we distinguish between in-degree (the number of incoming links) and out-degree (the number of outgoing links).

\item {\bf Eigenvector centrality:}
Defines a node’s score proportional to the sum of its neighbors’ scores, i.e.\ as the principal eigenvector of the adjacency matrix  (\citet{bonacich2007}). It highlights nodes connected to other well‐connected nodes, and in friendship‐paradox terms, identifies those whose friends are themselves highly central.

\item {\bf Walk counts:} For an integer \(\ell \ge 0\) we define the walk count vector
\(\br = A^\ell \mathbf{1}\); the entry \(r_i\) equals the number of walks of length
\(\ell\) starting at node \(i\). Although this is not used as a centrality measure but we include it as it forms a basic building block for other centralities. Walk-based centrality indices of this type appear in \citet{estrada2010generalized} and in related communicability measures \citet{estrada2009communicability}. More broadly, many authors developed a family of matrix-function walk-based indices, including subgraph/communicability-based centralities, odd/even variants and functional centralities, and resolvent-based analogues; see e.g. \citet{estrada2005subgraph, rodriguez2007functional, estrada2010network}. 

\item {\bf Katz centrality:} This extends degree centrality by counting all walks emanating from a node, exponentially down‐weighting longer walks via a parameter $\alpha<\frac{1}{\rho(A)}$ (\citet{katz1953new}). It captures both direct ties and indirect influence, but still emphasizes nearer neighbors more heavily.

\item {\bf PageRank centrality:}
A variant of eigenvector centrality that assumes a random‐walk with occasional “teleportation” (restart) to any node with probability $1-\beta$ (\citet{page1999pagerank}). PageRank reflects both the quantity and quality of friends, and mitigates sink effects by redistributing rank.

\end{enumerate}

We provide a separate definition of the generalized friendship paradox for centrality indices that satisfy certain properties, and we refer to this as the centrality paradox.
\begin{defin}\label{def:paradox}
{\bf Centrality paradox}: Given a finite connected graph $G=(V, E)$ with $|V|=n$. If node-wise centrality $\br=(r_1,\dots,r_n)^\top$, then define
\[
\overline{\mu}_{\br}
\;=\;
\frac{1}{n}\sum_{i=1}^n \Bigl(\,\frac{1}{d_i}\sum_{j=1}^n A_{ij} r_j\Bigr)
\quad\text{and}\quad
\mu_{\br}
\;=\;
\frac{1}{n}\sum_{i=1}^n r_i.
\]

We say that the centrality index $\br$ exhibits a \emph{(neighbour-averaged)
centrality paradox } if
\[
\overline{\mu}_{\br}\;\ge\;\mu_{\br}.
\]
 Recall, the $C$ matrix as defined in \eqref{eq:Cmatrix}, then the centrality paradox is equivalent to
$$
\sum_{i=1}^n \Bigl( \sum_{j=1}^n C_{ij} r_j - r_i\Bigr)\ge 0,
$$
or, equivalently, 
\begin{equation}\label{eq:cdef}
\scalar{\one}{C\br} \ge \scalar{\one}{\br}.
\end{equation}

\end{defin}

In subsection~\ref{subsec:pagerank} we show that an analogous inequality also holds
for PageRank on strongly connected directed graphs, with \(C\) defined from the
out-degrees.

\begin{theorem}[Centrality Paradox]
  \label{thm:main}
  Let $G=(V,E)$ be a connected, undirected graph on $n$ vertices, with adjacency matrix $A$, degree vector $\mathbf d$, and Perron eigenvalue~$\lambda_1$.  We consider the following centralities $\mathbf r$,
 \[
    \begin{aligned}
      &\text{Degree:}        &\mathbf r &= A\,\mathbf1,\\
      &\text{Walk‐count (order }\ell\text{):}   
                             &\mathbf r &= A^{\ell}\,\mathbf1,\\
      &\text{Eigenvector:}   &A\,\mathbf r &= \lambda_1\,\mathbf r,\\
      &\text{Katz:}          &\mathbf r &= (I - \alpha A)^{-1}\mathbf1,\quad 0<\alpha<1/\lambda_1,\\
      &\text{PageRank:}      &\mathbf r &= (1-\beta)\,C\,\mathbf r + \beta\tfrac{1}{n}\mathbf1, \qquad \beta\in(0,\, 1).
    \end{aligned}
  \]  
{Then in each case inequality \eqref{eq:cdef} holds, i.e., these centralities
exhibit the (neighbour-averaged) centrality paradox}. 
\end{theorem}


Some of the proofs depend crucially on the following variational formulation of the eigenvector corresponding to the Perron eigenvalue.
\begin{theorem}[\citet{Fiedler1985}]\label{thm:PFbilinear}
Let $P\in\R^{n\times n}$ be irreducible and nonnegative, with the Perron eigenvalue $\lambda>0$, right eigenvector $\bu>0$, and left eigenvector $\bv>0$ normalized so that $\bv^\top \bu=1$.  Then for any $\bx,\by>0$ with $\bx\circ\by=\bu\circ\bv$ (where $\circ $ is the Hadamard product $
\bx\circ\by =(x_1y_1,\ldots,x_ny_n)^\top$), 
\[
\by^\top P\,\bx \;\ge\; \bv^\top P\,\bu = \lambda,
\]
and equality occurs if and  only if  $\bx$ is a scalar multiple of $\bu$ (whence $\by$ is the corresponding inverse scalar multiple of $\bv$).  In particular, setting $\bx=\bv$ and $\by=\bu$ gives
\[
\bu^\top P\,\bv \;\ge\; \bv^\top P\,\bu = \lambda,
\]
with equality if and  only if  $\bu$ and $\bv$ are linearly dependent.
\end{theorem}

\section{Centrality Paradox for Specific Centralities}\label{sec:examples}

Below we show that for specific centralities the paradox is true. 

\subsection{Degree Centrality}
We now prove the paradox for degree centrality. Recall that
\(\br = \bd = A\mathbf{1}\).
  Then
\[
\langle \one,\,C\,\bd\rangle 
\;=\; 
\langle \one,\,D^{-1}A\,\bd\rangle
\;=\;\sum_{i=1}^n \frac{1}{d_i}\,\bigl(A\,\bd\bigr)_i
\;=\;\sum_{i=1}^n \frac{1}{d_i}\sum_{j=1}^n A_{ij}\,d_j,
\]
while 
\(\langle \one,\,\bd\rangle = \sum_{i=1}^n d_i.\)
Thus
\[
\langle \one,\,C\,\bd\rangle - \langle \one,\,\bd\rangle
\;=\;
\sum_{i=1}^n \frac{1}{d_i}\sum_{j=1}^n A_{ij} d_j
\;-\; \sum_{i=1}^n d_i 
\;=\; 
\frac{1}{2}\sum_{i=1}^n\sum_{j=1}^n A_{ij}\Bigl(\sqrt{\tfrac{d_j}{d_i}} - \sqrt{\tfrac{d_i}{d_j}}\Bigr)^2
\;\ge\;0.
\]
Hence 
\(\langle \one,\,C\,\bd\rangle \ge \langle \one,\,\bd\rangle,\)
with strict inequality unless all degrees are equal (i.e.\ \(G\) is regular). Instead of the symmetrization trick one can use the variational formulation of Theorem \ref{thm:PFbilinear} to show this result as well.

Since $C$ is irreducible, $1$ is its simple Perron eigenvalue with right eigenvector $\one$ and left eigenvector $\bw=\bd / \|\bd\|_1$.  By Theorem~\ref{thm:PFbilinear}, choosing $\bx=\bw$ and $\by=\one$ gives
\[
\langle \one,\,C\bw\rangle \;\ge\; \langle \bw,\,C\one\rangle=\langle \bw,\,\one\rangle=1,
\]
i.e.
\[
\sum_{i=1}^n \Bigl(\tfrac{1}{d_i\|\bd\|_1}\sum_{j=1}^n A_{ij}\,d_j\Bigr)
\;\ge\;
\sum_{i=1}^n \tfrac{d_i}{\|\bd\|_1}=1,
\]
or equivalently,
\[
\sum_{i=1}^n \frac{1}{d_i}\sum_{j=1}^n A_{ij} d_j \;\ge\; \sum_{i=1}^n d_i,
\]
recovering the classical friendship paradox for degree.  

\begin{remark}
    The proof of the classical friendship paradox using symmetrization is standard; we include it only
for completeness.
\end{remark}

\subsection{Eigenvector centrality}
We now show the paradox holds true for eigenvector centrality.  Recall, that it solves
\[
A\,\br = \lambda_1\,\br,\qquad \|\br\|_1=1,\quad \br>0.
\]

In eigenvector centrality a node with relatively few links can score highly if they are adjacent to other nodes with high eigenvector scores or highly central nodes. Formally, each node’s centrality is defined to be proportional to the sum of the centralities of its neighbors, which means that nodes positioned close to the most prominent individuals or tightly knit communities naturally attain higher eigenvector scores.

Since \(A\) is symmetric, \(\langle \br,\,A\,\one\rangle = \langle A\,\br,\,\one\rangle = \lambda_1\,\langle \br,\,\one\rangle = \lambda_1.\)
We need to show \eqref{eq:cdef} and hence we compute
\[
\langle \one,\,C\,\br\rangle 
\;=\; 
\langle \one,\,D^{-1}A\,\br\rangle
\;=\;\sum_{i=1}^n \frac{1}{d_i}\,(A\,\br)_i
\;=\;\sum_{i=1}^n \frac{\lambda_1\,r_i}{d_i}.
\]
Meanwhile \(\langle \one,\,\br\rangle = 1.\)  Thus we need
\[
\sum_{i=1}^n \frac{\lambda_1\,r_i}{d_i} \;\ge\; 1,
\quad\text{i.e.}\quad
\sum_{i=1}^n \frac{r_i}{d_i} \;\ge\; \frac{1}{\lambda_1}.
\]
As $\sum_{i=1}^n r_i=1$ by the weighted harmonic–arithmetic mean inequality (\citet{maze2009note}) we have
\[
\sum_{i=1}^n \frac{r_i}{d_i} 
\;\ge\; 
\frac{1}{\sum_{i=1}^n r_i\,d_i} 
\;=\; 
\frac{1}{\langle \br,\,\bd\rangle}
\;=\; 
\frac{1}{\langle \br,\,A\,\one\rangle}
\;=\; 
\frac{1}{\lambda_1}.
\]

Therefore 
$\langle \one,\,C\,\br\rangle \ge \langle \one,\,\br\rangle$, 
with equality if and  only if  all \(d_i\) coincide (i.e.\ \(G\) is regular).

\subsection{Walk Counts}
We now consider the walk counts \(\br = A^\ell \mathbf{1}\).
Then
\[
\langle \one,\,C\,\br\rangle 
\;=\; 
\langle \one,\,D^{-1}A\,A^\ell\,\one\rangle
\;=\;\sum_{i=1}^n \frac{1}{d_i}\,(A^{\ell+1}\one)_i,
\quad
\langle \one,\,\br\rangle = \sum_{i=1}^n (A^\ell\one)_i.
\]

{We use a simple inequality for symmetric nonnegative matrices. Let
\(W = (W_{ij})\) be symmetric and nonnegative and let \(x = (x_i)_{i=1}^n\) have
strictly positive entries. Then
\[
\sum_{i,j=1}^n x_i W_{ij} x_j^{-1}
  = \sum_{i=1}^n W_{ii}
    + \sum_{1 \le i < j \le n} W_{ij}
       \Bigl(\frac{x_i}{x_j} + \frac{x_j}{x_i}\Bigr)
  \ge \sum_{i=1}^n W_{ii}
    + 2 \sum_{1 \le i < j \le n} W_{ij}
  = \sum_{i,j=1}^n W_{ij},
\]
where we used that \(t + t^{-1} \ge 2\) for all \(t>0\). Applying this with
\(W = A^\ell\) and \(x_i = 1/d_i\) gives
\[
\sum_{i,j=1}^n \frac{1}{d_i} W_{ij} d_j
  \ge \sum_{i,j=1}^n W_{ij}=\one^\top W\,\one.
\]
}
Hence
\[
\langle \mathbf{1}, C \br\rangle
 = \sum_{i=1}^n \frac{1}{d_i} (A^{\ell+1}\mathbf{1})_i
 = \sum_{i,j=1}^n \frac{1}{d_i} W_{ij} d_j
 \ge \sum_{i,j=1}^n W_{ij}
 = \langle \mathbf{1}, A^\ell \mathbf{1}\rangle
 = \langle \mathbf{1}, \br\rangle.
\]
For \(\ell=0\), we have \(W = I\) and the inequality holds with equality.


\subsection{Katz Centrality}
We show that Katz centrality follows the paradox. Let \(\lambda_1\) be the Perron eigenvalue of \(A\) from Section~\ref{sec:paradox-def} and choose
\(0<\alpha<1/\lambda_1\). Here $\br$ satisfies
\[
\br = (I - \alpha A)^{-1}\,\one \;=\;\sum_{\ell=0}^\infty \alpha^\ell\,A^\ell\,\one,
\]
which converges entrywise because \(\alpha\,\lambda_1<1\).   Note
\[
\langle \one,\,C\,\br\rangle 
\;=\; 
\langle \one,\,D^{-1}A\,\bigl((I-\alpha A)^{-1}\one\bigr)\rangle
\;=\; \sum_{\ell=0}^\infty \alpha^\ell 
\sum_{i=1}^n \frac{(A^{\ell+1}\one)_i}{d_i},
\]
whereas
\(\langle \one,\,\br\rangle 
=\sum_{\ell=0}^\infty \alpha^\ell \sum_{i=1}^n (A^\ell\one)_i.\)
Therefore
\[
\langle \one,\,C\,\br\rangle - \langle \one,\,\br\rangle
= \sum_{\ell=0}^\infty \alpha^\ell 
\Bigl[\sum_{i=1}^n \tfrac{(A^{\ell+1}\one)_i}{d_i} - \sum_{i=1}^n (A^\ell\one)_i\Bigr].
\]
But by the argument above for walk counts, one has
\[
\sum_{i=1}^n \tfrac{(A^{\ell+1}\one)_i}{d_i} 
\;\ge\; 
\sum_{i=1}^n (A^\ell\one)_i,
\]
with equality if and  only if \(G\) is regular.  Since \(\alpha^\ell>0\), each summand is nonnegative and at least one is strictly positive when \(G\) is non-regular.  Therefore
\(\langle \one,\,C\,\br\rangle \ge \langle \one,\,\br\rangle.\)

\subsection{PageRank Centrality (Directed Setting)}\label{subsec:pagerank}

PageRank refines eigenvector centrality by dampening the effect that a highly central node has when it links to many others. It evaluates each node’s importance based on both the quantity and quality of its inbound links, in particular,  a link from a modestly connected page carries more weight than one from an out-degree “hub.” PageRank balances three factors: the number of incoming links, each linker’s tendency to distribute its opinion across its own outgoing edges, and the linking page’s own centrality. {By incorporating this topology-driven feedback with a small probability of “teleporting” to any node, PageRank defines a stationary distribution for the modified random walk and is widely used as a centrality index in directed networks.}
In this subsection, we allow \(G=(V,E)\) to be a \emph{strongly connected directed} graph on \(n\) nodes.  Write $A$ to be adjacency matrix, that is, 
\[
A_{ij}=1 \text{ if there is a directed edge } i\to j, \text{ and } 0 \text{ otherwise}.
\]
Define the out‐degree vector
\[
d_i^{\mathrm{out}} \;=\; \sum_{j=1}^n A_{ij}, \qquad 
D_{\mathrm{out}}=\mathrm{diag}\bigl(d_1^{\mathrm{out}},\dots,d_n^{\mathrm{out}}\bigr).
\]
Since \(G\) is strongly connected,  \(d_i^{\mathrm{out}}>0\) for all $i$.  We reformulate the transition matrix $C$ as follows
\[
C \;=\; D_{\mathrm{out}}^{-1}A,
\quad
C_{ij} = 
\begin{cases}
\frac{1}{d_i^{\mathrm{out}}}, & i\to j,\\
0, & \text{otherwise.}
\end{cases}
\]
Note that \(C\) is irreducible and \(C\,\one = \one\). Fix a teleportation parameter \(\beta\in(0,1)\) and define the PageRank matrix
\[
P \;=\; (1-\beta)\,C +  \beta\,\frac{1}{n} \one \one^\top .
\]
Since \(C\) is irreducible and \(\mathbf v>0\), \(P\) is also irreducible and {row‐stochastic}.  Hence the Perron eigenvalue of \(P\) is \(\lambda=1\). The PageRank centrality $\br$ is defined as the unique  normalized ($\langle \one,\br \rangle=1$) 
solution of 
    \begin{equation}\label{def:pagerank}
   \br^{\top} P=\br^{\top}.
    \end{equation}

We would like to prove the directed‐PageRank paradox, that is,
\begin{equation}\label{eq:inequalityPR}
\overline{\mu}_{\br}
\;=\;
\frac{1}{n}\sum_{i=1}^n \Bigl(\tfrac{1}{d_i^{\mathrm{out}}}\sum_{\,j=1}^n A_{ij} r_j\Bigr)
\;\ge\;
\frac{1}{n}\sum_{i=1}^n r_i 
\;=\; \mu_{\br},
\end{equation} 
equivalently, $ \langle \one,\,C\,\br\rangle \;\ge\; \langle \one,\,\br\rangle=1$. {This is the directed analogue of Definition~\ref{def:paradox} with the same neighbour-averaged
quantity \(\langle \mathbf{1}, C \br\rangle\), but with \(C\) built from the
directed out-degrees.}

\medskip

Since \(P\) is irreducible and row‐stochastic, we can apply Theorem~\ref{thm:PFbilinear}
with $\br$ and $\one$ being the left and right eigenvectors, respectively: 
\begin{equation}\label{eq:Peq}
\one^\top\,P\,\br 
\;\ge\; \br^\top\,P\,\one 
\;=\; \langle \br,\one\rangle 
\;=\; 1,
\end{equation}

Using the definition of $P$ one has \[
\one^\top\,P\,\br 
= \one^\top\Bigl[(1-\beta)\,C + \frac{\beta}{n}\,\one \one^\top\Bigr]\br
= (1-\beta)\,\one^\top\,C\,\br \;+\; \frac{\beta}{n}\,\one^\top\one (\one^\top\,\br)
= (1-\beta)\,\one^\top\,C\,\br \;+\; \beta.
\]
Hence from \eqref{eq:Peq} it follows
\[
(1-\beta)\,\one^\top\,C\,\br + \beta \;\ge\; 1 
\quad\Longleftrightarrow\quad
\one^\top\,C\,\br \;\ge\; 1 \;=\; \one^\top\,\br.
\]
Hence this shows \eqref{eq:inequalityPR}.

\begin{remark}Here we average over outgoing neighbours: we first choose a node uniformly at
random and then pick one of its outgoing neighbours uniformly at random. This
matches the usual random-walk interpretation of PageRank, for which
$C = D_{\text{out}}^{-1}A$ is row-stochastic and $\langle\mathbf{1},Cr\rangle$
is the average centrality seen along these outgoing edges. Other choices in the
directed setting, such as averaging over in-neighbours using $D_{\text{in}}^{-1}A^\top$,
lead to different but related formulations; see the discussion in Section~5.
\end{remark}

\section{Comparison of centrality paradox}\label{sec:comparison}
The generalized friendship paradox has appeared in two closely related but distinct forms in the literature. Our paper uses a node-sampled version, which leads to $\bar{\mu}_r=\frac{1}{n} \mathbf{1}^{\top} C r$. Earlier works \cite{higham2019centrality} use an edge-sampled version (pick an edge uniformly, then pick an endpoint), which yields $\tilde{\mu}_r=\frac{d^{\top} r}{d^{\top} 1}$. Since these correspond to different empirical questions, we briefly compare them; in general neither dominates, so the two formulations are complementary rather than interchangeable.

In the discussion above we have focused on the “neighbour‐averaging” version of the centrality paradox, namely
\[
\bar\mu_r 
\;=\;\frac{1}{n}\sum_{i=1}^n\frac{1}{d_i}\sum_{j=1}^n A_{ij} r_j,
\]
which compares each node’s own score to the average score of its neighbours. An alternative, equally natural way to weight friends is by sampling edges uniformly at random and then looking at the centrality of the endpoint reached. This leads to the \emph{edge‐weighted average}
\[
\tilde\mu_r
\;=\;\frac{\sum_{i=1}^n d_i\,r_i}{\sum_{k=1}^n d_k}
\;=\;\frac{\br^\top \bd}{\one^\top \bd},
\]
where \(\bd=(d_1,\dots,d_n)^T\) is the degree vector. In this section we compare $\bar \mu_r$ and $\tilde \mu_r$. Note that they agree when all the degrees $d_i$ are same. The centrality paradox for some of the centrality indicess in terms of $\tilde \mu_r$ was studied in \citet{higham2019centrality}.
{In this section we
compare these two natural friends’ averages, \(\bar\mu_r\) and \(\tilde\mu_r\),
to understand how the formulation relates to each other. }

Recall the two “friend‐average” quantities for a centrality vector \(\br\):
\[
\tilde \mu_{\br}
\;=\;
\frac{\br^\top \bd}{\one^\top \bd}
\;=\;
\frac{\sum_{j=1}^n r_j\,d_j}{\sum_{k=1}^n d_k},
\qquad
\bar \mu_{\br}
\;=\;
\frac{1}{n}\sum_{i=1}^n \frac{1}{d_i}\sum_{j\sim i} r_j.
\]
We can rewrite \(\bar \mu_{\br}\) by swapping the order of summation:
\[
\bar \mu_{\br}
\;=\;
\frac{1}{n}\sum_{i=1}^n \frac{1}{d_i}\sum_{j\sim i} r_j
\;=\;
\frac{1}{n}\sum_{j=1}^n r_j \sum_{\,i:\,i\sim j} \frac{1}{d_i}.
\]
Thus
\[
\bar \mu_{\br}
\;=\;
\frac{1}{n}\sum_{j=1}^n r_j\,a_j,
\quad
\text{where}\quad
a_j = \sum_{\,i:\,i\sim j}\frac{1}{d_i}.
\]
Meanwhile,
\[
\tilde \mu_{\br}
\;=\;
\frac{\sum_{j=1}^n r_j\,d_j}{\sum_{k=1}^n d_k}
\;=\;
\sum_{j=1}^n r_j\,b_j,
\quad
\text{where}\quad
b_j = \frac{d_j}{\sum_{k=1}^n d_k}.
\]
Therefore,
\[
\bar \mu_{\br} - \tilde \mu_{\br}
\;=\;
\sum_{j=1}^n r_j\Bigl(\tfrac{1}{n}a_j - b_j\Bigr).
\]
In general there is no fixed inequality between \(\bar \mu_{\br}\) and \(\tilde \mu_{\br}\) without further assumptions on \(\br\) or \(G\); they coincide when \(G\) is regular (all \(d_i\) equal), in which case \(\bar \mu_{\br}=\tilde \mu_{\br}=\mu_{\br}\). These two quantities in general need not coincide.

{\bf Example:} Consider the degree centrality and the star graph on \(n\) nodes. A simple calculation shows
  \[
    \tilde\mu_{d}
      \;=\; \frac{(n-1)^2 + (n-1)\cdot1^2}{2(n-1)} \;=\; \frac{n}{2}, 
    \quad
    \bar\mu_{d}
      \;=\; \frac{1 + (n-1)^2}{n},
  \]
  so \(\bar\mu_{d}>\tilde\mu_{d}\) for \(n\ge3\).

  In general connected graphs, it is unclear if there is an universal ordering between \(\bar\mu_{d}\) and \(\tilde\mu_{d}\).

  Now instead of degree centrality, if one considers the eigenvector centrality, then one can show in certain cases, $\tilde \mu_{r}>\bar \mu_r$.

{\bf Example:} 
Consider  a line graph on $6$ vertices:
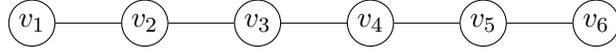
\begin{center}
\begin{tikzpicture}[scale=1, every node/.style={circle, draw, fill=white, inner sep=2pt}]
  \node (v1) at (0,0)  {$v_1$};
  \node (v2) at (1.5,0) {$v_2$};
  \node (v3) at (3,0)   {$v_3$};
  \node (v4) at (4.5,0) {$v_4$};
  \node (v5) at (6,0)   {$v_5$};
  \node (v6) at (7.5,0) {$v_6$};

  \draw (v1) -- (v2);
  \draw (v2) -- (v3);
  \draw (v3) -- (v4);
  \draw (v4) -- (v5);
  \draw (v5) -- (v6);
\end{tikzpicture}
\end{center}
The adjacency matrix is given by
\[
A \;=\;
\begin{pmatrix}
0 & 1 & 0 & 0 & 0 & 0 \\[6pt]
1 & 0 & 1 & 0 & 0 & 0 \\[6pt]
0 & 1 & 0 & 1 & 0 & 0 \\[6pt]
0 & 0 & 1 & 0 & 1 & 0 \\[6pt]
0 & 0 & 0 & 1 & 0 & 1 \\[6pt]
0 & 0 & 0 & 0 & 1 & 0
\end{pmatrix}
\]
The characteristic polynomial is given by $\lambda^6-5\lambda^4+6\lambda -1$ and the eigenvalues are 
$$
\lambda_k=2\cos\left(\frac {\pi k}{7}\right),\quad k=1,2,\ldots,6,
$$
with the corresponding right eigenvectors 
\[
\mathbf{v}^{(k)} = 
\begin{bmatrix}
\sin\left( \dfrac{k \pi}{7} \right),
\sin\left( \dfrac{2k \pi}{7} \right),
\sin\left( \dfrac{3k \pi}{7} \right),
\sin\left( \dfrac{4k \pi}{7} \right),
\sin\left( \dfrac{5k \pi}{7} \right),
\sin\left( \dfrac{6k \pi}{7} \right) 
\end{bmatrix}^{\top}.
\]
The Perron eigenvalue is $\lambda=\lambda_1=2\cos \left(\frac {\pi }{7}\right)\approx 1.80194$ with the normalized eigenvector
$$
{\br}=[0.0990, \quad
0.1785,\quad
0.2224,\quad
0.2224, \quad
0.1785, \quad
0.0990]^{\top}.
$$
If we consider the degree centrality  $d=[1,2,2,2,2,1]^{\top}$, then
$$\aligned
\bar \mu_{d}&=\frac { \frac 11\cdot 2+ \frac 12(1+2)+\frac 12(2+2)+\frac 12(2+2)+
\frac 12(2+1)+\frac 1{1}\cdot2}{6}=\frac {11}{6}\\
\tilde \mu_{d}&=\frac {1+4+4+4+4+1}{1+2+2+2+2+1}=\frac{14}{10}.
\endaligned
$$
We see that $\bar \mu_{d}>\tilde  \mu_{d}$.
However, if we consider eigenvector centrality $\br $ and $\bd^{(2)}=A\bd$ we have
$$\begin{gathered} \bar \mu_{\br}\approx 0.1799,\quad \bar \mu_{\br}\approx 0.1822,\\
 \bar \mu_{\bd^{(2)}}\approx 5.8333,\quad \bar \mu_{\bd^{(2)}}\approx 5.875,
 \end{gathered}
$$
i.e., the opposite inequalities hold: $\bar \mu_{\br}<\tilde \mu_{\br}$ and $ \bar \mu_{\bd^{(2)}}< 
\tilde \mu_{\bd^{(2)}}$.

\section{Discussion: Open directions}\label{sec:discussion}
We now end with some discussions on the open directions. We have shown that, in every non‐regular connected graph, the average centrality of a node’s neighbors strictly exceeds that node’s own score for a wide class of measures such as degree, eigenvector, walk counts, Katz and PageRank centralities. This unified ``centrality paradox" thus holds across both local and global functionals. Here are some of the crucial open areas which immediately emerge from the findings of this article.

\begin{enumerate}
    \item Not all popular measures fit our current framework. Closeness and harmonic centrality depend on shortest-path distances rather than powers or resolvents of the adjacency matrix, so our bilinear–inequality arguments (which hinge on quadratic forms and spectral identities) do not immediately apply. Determining whether the neighbor-averaging paradox holds for these distance-based measures remains open and likely requires novel comparisons of inverse-distance sums to their global averages.

    \item In case of sparse random graphs such as Erd\H{o}s-R\'enyi, configuration model and preferential attachment model, \citet{HHP1, HHP2, HHNP} analyze the “friendship bias” for degree centrality. Here “friendship bias” refers to the random difference between the average neighbour degree and the node’s own degree.
 One can similarly define the bias for centrality index by
$$\Delta_i^{(r)}=\frac{1}{d_i} \sum_{j \sim i} r_j-r_i.$$
Now one can define the empirical measure $$\mu_{\mathrm{bias},n}=\frac{1}{n}\sum_{i=1}^n\delta_{\Delta_i^{(r)}},$$ where $\delta$ is the dirac-delta measure. This measure contains non-trivial information about the paradox. In the case of classical friendship paradox with degree centrality, using the tools of local weak convergence (\citet{van2024random}) one can show the above measure converges and can be expressed explicitly in terms of some random trees.  In contrast, eigenvector and Katz centralities are inherently global: their scores at every node reflect the full network. One cannot simply replace the finite graph by its local tree limit and still capture these spectral quantities. Although one can be hopeful as these methods have been used for various centrality indices to study their properties, for example \citet{garavaglia2020local} study the local weak convergence results for PageRank centrality. See also \citet{van2024connectivity} for other applications of local weak convergence and centrality indices. So the open direction would be to study the centrality indices mentioned in this article and prove the convergence of $\mu_{\mathrm{bias},n}$.

\item A promising avenue is to blend our paradox inequalities with spectral analyses
on sparse, tree-like graphs, such as the work of \citet{bordenave2010resolvent},
which characterizes the limiting behaviour of eigenvalues of large random
adjacency or non-backtracking matrices via {Galton--Watson trees, that is, random rooted trees generated by a branching
process and arising as local weak limits of many sparse random graphs. If one
can analyse the eigenvectors on such explicit random trees, it may become
possible to derive precise limit laws for the centrality paradox for
eigenvectors.} Also in sparse random graphs there is a subtle interplay of localization and delocalization of eigenvectors and it would be interesting to find connections of the eigenvector centrality paradox and such physical phenomenon.

    \item In directed networks, there is also a modelling choice in how to define a
“friend of a node.” Our directed PageRank result in Section~\ref{subsec:pagerank} uses
out-neighbours, in line with the underlying random walk that moves along
outgoing edges and with the row-stochastic matrix $C = D_{\text{out}}^{-1}A$.
One could instead average over in-neighbours, or over a symmetrised version of
the directed graph, which would lead to different centrality-paradox
formulations; analysing these variants remains an open question.
\end{enumerate}

\section*{Acknowledgement} 
The authors thank Frank den Hollander and Nelly Litvak for the discussion around the topic. RSH was supported by the Netherlands Organisation for Scientific Research (NWO)
through Gravitation-grant NETWORKS-024.002.003.
We thank the Editor and two anonymous referees for very useful suggestions.

\bibliographystyle{abbrvnat}
\bibliography{reference.bib}

\begin{thebibliography}{30}
\providecommand{\natexlab}[1]{#1}
\providecommand{\url}[1]{\texttt{#1}}
\expandafter\ifx\csname urlstyle\endcsname\relax
  \providecommand{\doi}[1]{doi: #1}\else
  \providecommand{\doi}{doi: \begingroup \urlstyle{rm}\Url}\fi

\bibitem[Avrachenkov and Dreveton(2022)]{avrachenkov2022statistical}
K.~Avrachenkov and M.~Dreveton.
\newblock \emph{Statistical Analysis of Networks}.
\newblock Now Publishers, 2022.

\bibitem[Bonacich(2007)]{bonacich2007}
P.~Bonacich.
\newblock Some unique properties of eigenvector centrality.
\newblock \emph{Social networks}, 29\penalty0 (4):\penalty0 555--564, 2007.

\bibitem[Bordenave and Lelarge(2010)]{bordenave2010resolvent}
C.~Bordenave and M.~Lelarge.
\newblock Resolvent of large random graphs.
\newblock \emph{Random Structures \& Algorithms}, 37\penalty0 (3):\penalty0
  332--352, 2010.

\bibitem[Cantwell et~al.(2021)Cantwell, Kirkley, and
  Newman]{cantwell2021friendship}
G.~T. Cantwell, A.~Kirkley, and M.~E. Newman.
\newblock The friendship paradox in real and model networks.
\newblock \emph{Journal of Complex Networks}, 9\penalty0 (2):\penalty0 cnab011,
  2021.

\bibitem[Christakis and Fowler(2008)]{christakis2008collective}
N.~A. Christakis and J.~H. Fowler.
\newblock The collective dynamics of smoking in a large social network.
\newblock \emph{New England journal of medicine}, 358\penalty0 (21):\penalty0
  2249--2258, 2008.

\bibitem[Eom and Jo(2014)]{eom2014generalized}
Y.-H. Eom and H.-H. Jo.
\newblock Generalized friendship paradox in complex networks: The case of
  scientific collaboration.
\newblock \emph{Scientific reports}, 4\penalty0 (1):\penalty0 4603, 2014.

\bibitem[Estrada(2010)]{estrada2010generalized}
E.~Estrada.
\newblock Generalized walks-based centrality measures for complex biological
  networks.
\newblock \emph{Journal of theoretical biology}, 263\penalty0 (4):\penalty0
  556--565, 2010.

\bibitem[Estrada and Higham(2010)]{estrada2010network}
E.~Estrada and D.~J. Higham.
\newblock Network properties revealed through matrix functions.
\newblock \emph{SIAM review}, 52\penalty0 (4):\penalty0 696--714, 2010.

\bibitem[Estrada and Rodriguez-Velazquez(2005)]{estrada2005subgraph}
E.~Estrada and J.~A. Rodriguez-Velazquez.
\newblock Subgraph centrality in complex networks.
\newblock \emph{Physical Review E—Statistical, Nonlinear, and Soft Matter
  Physics}, 71\penalty0 (5):\penalty0 056103, 2005.

\bibitem[Estrada et~al.(2009)Estrada, Higham, and
  Hatano]{estrada2009communicability}
E.~Estrada, D.~J. Higham, and N.~Hatano.
\newblock Communicability betweenness in complex networks.
\newblock \emph{Physica A: Statistical Mechanics and its Applications},
  388\penalty0 (5):\penalty0 764--774, 2009.

\bibitem[Feld(1991)]{Feld1991}
S.~L. Feld.
\newblock Why your friends have more friends than you do.
\newblock \emph{American journal of sociology}, 96\penalty0 (6):\penalty0
  1464--1477, 1991.

\bibitem[Fiedler et~al.(1985)Fiedler, Johnson, Markham, and
  Neumann]{Fiedler1985}
M.~Fiedler, C.~R. Johnson, T.~L. Markham, and M.~Neumann.
\newblock A trace inequality for m-matrices and the symmetrizability of a real
  matrix by a positive diagonal matrix.
\newblock \emph{Linear Algebra and its Applications}, 71:\penalty0 81--94,
  1985.

\bibitem[Garavaglia et~al.(2020)Garavaglia, van~der Hofstad, and
  Litvak]{garavaglia2020local}
A.~Garavaglia, R.~van~der Hofstad, and N.~Litvak.
\newblock Local weak convergence for page rank.
\newblock \emph{The Annals of Applied Probability}, 30\penalty0 (1):\penalty0
  40--79, 2020.

\bibitem[Hazra et~al.(2025{\natexlab{a}})Hazra, den Hollander, and
  Parvaneh]{HHP1}
R.~S. Hazra, F.~den Hollander, and A.~Parvaneh.
\newblock The friendship paradox for sparse random graphs.
\newblock \emph{Probability Theory and Related Fields}, pages 1--23,
  2025{\natexlab{a}}.

\bibitem[Hazra et~al.(2025{\natexlab{b}})Hazra, Hollander, Litvak, and
  Parvaneh]{HHNP}
R.~S. Hazra, F.~d. Hollander, N.~Litvak, and A.~Parvaneh.
\newblock The friendship paradox for trees.
\newblock \emph{arXiv preprint arXiv:2505.21774}, 2025{\natexlab{b}}.

\bibitem[Hazra et~al.(2025{\natexlab{c}})Hazra, Hollander, and Parvaneh]{HHP2}
R.~S. Hazra, F.~d. Hollander, and A.~Parvaneh.
\newblock The multi-level friendship paradox for sparse random graphs.
\newblock \emph{arXiv preprint arXiv:2502.17724}, 2025{\natexlab{c}}.

\bibitem[Higham(2019)]{higham2019centrality}
D.~J. Higham.
\newblock Centrality-friendship paradoxes: when our friends are more important
  than us.
\newblock \emph{Journal of Complex Networks}, 7\penalty0 (4):\penalty0
  515--528, 2019.

\bibitem[Hodas et~al.(2013)Hodas, Kooti, and Lerman]{hodas2013friendship}
N.~Hodas, F.~Kooti, and K.~Lerman.
\newblock Friendship paradox redux: Your friends are more interesting than you.
\newblock In \emph{Proceedings of the International AAAI Conference on Web and
  Social Media}, volume~7, pages 225--233, 2013.

\bibitem[Jackson(2019)]{jackson2019friendship}
M.~O. Jackson.
\newblock The friendship paradox and systematic biases in perceptions and
  social norms.
\newblock \emph{Journal of political economy}, 127\penalty0 (2):\penalty0
  777--818, 2019.

\bibitem[Katz(1953)]{katz1953new}
L.~Katz.
\newblock A new status index derived from sociometric analysis.
\newblock \emph{Psychometrika}, 18\penalty0 (1):\penalty0 39--43, 1953.

\bibitem[Kumar et~al.(2024)Kumar, Krackhardt, and Feld]{kumar2024friendship}
V.~Kumar, D.~Krackhardt, and S.~Feld.
\newblock On the friendship paradox and inversity: A network property with
  applications to privacy-sensitive network interventions.
\newblock \emph{Proceedings of the National Academy of Sciences}, 121\penalty0
  (30):\penalty0 e2306412121, 2024.

\bibitem[Maze and Wagner(2009)]{maze2009note}
G.~Maze and U.~Wagner.
\newblock A note on the weighted harmonic-geometric-arithmetic means
  inequalities.
\newblock \emph{arXiv preprint arXiv:0910.0948}, 2009.

\bibitem[Page et~al.(1999)Page, Brin, Motwani, and Winograd]{page1999pagerank}
L.~Page, S.~Brin, R.~Motwani, and T.~Winograd.
\newblock The pagerank citation ranking: Bringing order to the web.
\newblock Technical report, Stanford infolab, 1999.

\bibitem[Rodr{\'\i}guez et~al.(2007)Rodr{\'\i}guez, Estrada, and
  Guti{\'e}rrez]{rodriguez2007functional}
J.~A. Rodr{\'\i}guez, E.~Estrada, and A.~Guti{\'e}rrez.
\newblock Functional centrality in graphs.
\newblock \emph{Linear and Multilinear Algebra}, 55\penalty0 (3):\penalty0
  293--302, 2007.

\bibitem[Saxena and Iyengar(2020)]{saxena2020centrality}
A.~Saxena and S.~Iyengar.
\newblock Centrality measures in complex networks: A survey.
\newblock \emph{arXiv preprint arXiv:2011.07190}, 2020.

\bibitem[Shvydun(2025)]{shvydun2025zoo}
S.~Shvydun.
\newblock Zoo of centralities: Encyclopedia of node metrics in complex
  networks.
\newblock \emph{arXiv preprint arXiv:2511.05122}, 2025.

\bibitem[van~der Hofstad(2024)]{van2024random}
R.~van~der Hofstad.
\newblock \emph{Random graphs and complex networks}, volume~2.
\newblock Cambridge university press, 2024.

\bibitem[van~der Hofstad and Pandey(2024)]{van2024connectivity}
R.~van~der Hofstad and M.~Pandey.
\newblock Connectivity of random graphs after centrality-based vertex removal.
\newblock \emph{Journal of Applied Probability}, 61\penalty0 (3):\penalty0
  967--998, 2024.

\bibitem[Van~Mieghem(2014)]{Piet:book}
P.~Van~Mieghem.
\newblock \emph{Performance analysis of complex networks and systems}.
\newblock Cambridge University Press, 2014.

\bibitem[Wilson(2010)]{wilson2010using}
M.~Wilson.
\newblock Using the friendship paradox to sample a social network.
\newblock \emph{Physics today}, 63\penalty0 (11):\penalty0 15--16, 2010.

\end{thebibliography}

\end{document}